# Nature and Energy Source of the Strong Waveforms Recorded during the 2008 Wenchuan Earthquake

Xiaoping Mao[*1], Xueqiang Zhang[2], Yuche Su[1], Ke Mao[1], Pengyu Lu[1], and Fei Zhang[1]

[1]School of Energy Resource, China University of Geoscience, Chengfu Road 20, Haidian District, Beijing, China, 100086, ORCID: 0000-0002-2284-8833

[2]School of Earth Exploration and Information Technology, China University of Geosciences (Wuhan), Lumuo Road, Hongshan District, Wuhan, China, 430074

Corresponding author: Xiaoping Mao (maoxp9@163.com)

**Key Points:**

The energy of strong earthquake may come from fluid pressure release - fluid explosion in high porosity reservoir in sedimentary strata.

The strong amplitude waveform after the initial motion of the earthquake is not S-wave, that is, the possibility of the fluid explosion cannot be ruled out.

The phenomenon of electric field and ground explosion during the Wenchuan earthquake confirmed the possibility of such fluid activity.

Abstract

Earthquakes are indeed triggered by fault dislocations, but whether this process alone can produce the actual earthquake energy released by the mainshock has long been questioned. Therefore, exploring the true source of energy that causes earthquakes after the first motion is necessary. Based on analyses of the waveforms and ray paths at seismic stations close to the epicenter, it is considered that strong earthquake vibrations may not be caused by S-waves. It is also proposed that the reservoirs in sedimentary strata contain large amounts of high-pressure fluids, whose pressures can be released under certain conditions; this release of pressure may be an important component of the main earthquake energy. When a natural fault ruptures and penetrates a reservoir with a large area, the elastic energy produced by the release of pressure can reach the energy released by an earthquake of magnitude 8.0. Artificial engineering activities can lead to small-scale fluid pressure release phenomena, such as blowouts during drilling and earthquakes induced by hydraulic fracturing. Much direct and indirect evidence, such as the characteristics of seismic waves in the time and frequency domains recorded during the Wenchuan earthquake, explosion phenomena observed on the ground and cores obtained by scientific drilling, indicates the possibility of such energy release. We propose that seismicity can be divided into three stages: the microfracturing stage, in which there is fluid activity and can produce an electrokinetic effect; the significant fracturing stage after the initial movement; and the strong earthquake stage caused by fluid pressure release.

**Plain Language Summary**

At present, fault dislocation is considered to be the main cause and energy source of earthquakes, and it is believed that the strong earthquakes after the initial motion are mainly caused by S wave. Because the elastic energy released by fault dislocation alone is not enough to produce strong earthquake, this paper proposes another possibility of the source of strong earthquake energy: fluid pressure release in high porosity reservoir in sedimentary strata, fluid explosion, which can release the energy generated by earthquake of magnitude 8.

Therefore, based on the waveform data of seismic stations close to the epicenter of the actual Wenchuan earthquake, and through the ray path analysis and time-frequency analysis of seismic wave propagation, this paper concludes that the strong amplitude waveform after the initial motion of the earthquake is not S wave. Therefore, the possibility of P wave generated by fluid explosion can not be ruled out during the earthquake process.

A large number of direct or indirect evidences have confirmed the possibility of this fluid activity, such as the abnormal electric field phenomenon, the fluid activity phenomenon in the drilling core and the surface explosion phenomenon during the Wenchuan earthquake.

## 1 Introduction

Natural earthquakes are thought to be caused by tectonic movements (Brace & Byerlee, 1966). The 2008 Wenchuan Ms 7.9 earthquake is no exception and is considered to have been caused by tectonic movements, namely, shortening of the crust due to the collision of the Indian and Eurasian plates (Hubbard & Shaw, 2009). The methods for determining the seismic phase, studying the characteristics of earthquakes and locating earthquakes by using the initial motion of earthquakes are quite mature, whereas little research has been conducted on the strong earthquake waveforms after the initial motion. In most cases, strong waveforms are considered to be caused by S-waves (Honda, 1962). However, some scholars have raised doubts that the energy from tectonic movements alone, such as fault dislocations, is enough to generate the energy released by an earthquake. Gomberg et al. (2004) speculated that the stress change that triggers the earthquake is usually smaller than the magnitude of the stress released by the earthquake itself. Gilat and Vol (2005) calculated that the maximum strain energy of $600*100*20$ km$^3$ high-quality steel is $2.9\times10^{17}$ J, which is equivalent to a magnitude 8.4 earthquake, which is very extreme. Li et al. (2005) analyzed the distributions and characteristics of mining-induced seismicity over 120 coal mines and noncoal mines in mainland China from 1954 to 2005; among them, 47 produced earthquakes with recorded magnitudes, most of which were approximately Ms 2.8 (Figure 1). These were typical earthquakes with only fault dislocations, and the magnitudes were very small.

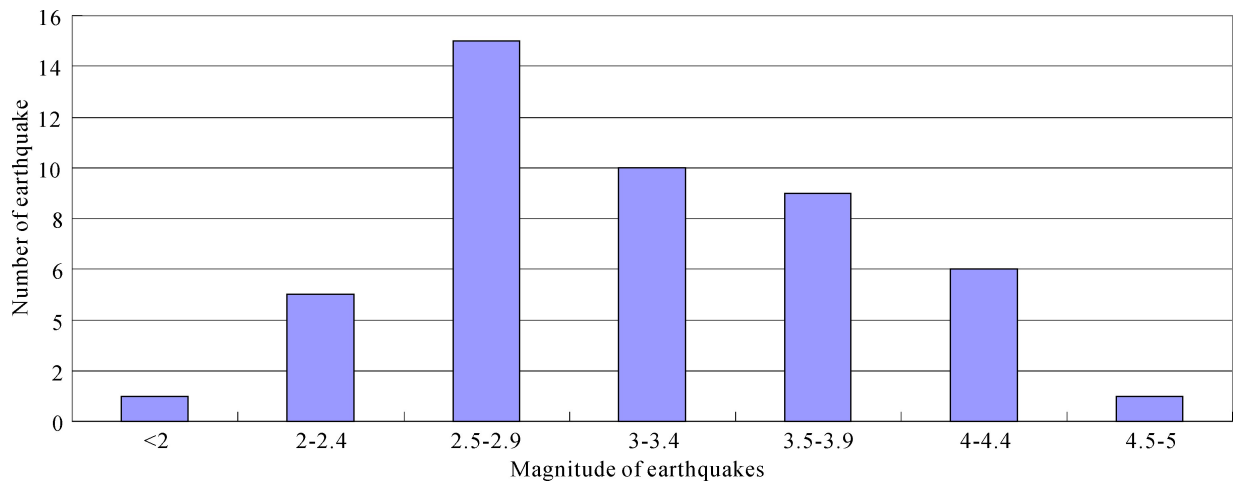

**Figure 1**. Distribution of the magnitudes of earthquakes caused by mining.

Certainly, earthquakes are triggered by tectonic movement along faults (Byerlee, 1970; Sokos et al., 2020), and the cause of serious damage to surface buildings may be the participation of fluid from the Earth's interior.

A number of studies have focused on quantifying the complex relationship between earthquakes and fluids. Many scholars attribute these phenomena of fluid activity to deep fluids, which can directly affect earthquakes. Zhao et al. (1996) proposed that the 1995 Kobe Ms 7.2 earthquake was the result of deep fluid accumulation at a depth of 16 km beneath the earthquake's epicenter. Kurz et al. (2004) suggested that an earthquake swarm in the European plate may have been triggered by deep fluid activity. Reyners et al. (2007) indicated that deep fluid activity may be a major cause of crustal seismic activity. Liu et al. (1996) inferred that the vertical force produced by the expansion of magma upwelling were the driving forces behind the Tangshan earthquake.

Many scholars even maintain that deep fluids play a leading role in earthquakes. Yue (2014) pointed out that the Wenchuan earthquake was caused by exploding of methane from the mantle. Liang (2017) concluded that the Wenchuan earthquake was a series of cryptoexplosions. Du et al. (2008) noted that the fluids in the core and lower mantle continuously escape upward toward the surface; in doing so, those fluids accumulate at different depths and may ultimately cause cryptoexplosions. Jamtveit et al. (2018) suggested that fluids in the lower crust drive metamorphism and structural transformation, leading to a significant decrease in lithospheric strength. Mandal (2019) attributed the occurrence of continuous earthquakes in the Kachchh rift belt in Gujarat to the release of $CO_2$ during the crystallization of carbonate melt. Heinicke et al. (2019) found that such $CO_2$ emissions are prevalent in the crust, leading to local weakening and slip within fault zones.

In this study, we explore the possibility that cryptoexplosions are triggered by these fluids.

## 2 The Possibility of Cryptoexplosions Caused by Deep Fluids

"Deep fluids" from the mantle exist in the metamorphic/crystalline basement (Figure 2). However, whether deep fluid can produce cryptoexplosions during earthquakes is the core of this paper. There are two types of petroleum reservoirs in sedimentary rocks: conventional reservoirs with high porosity and shales with low porosity and low permeability. To explain the possibility of cryptoexplosions produced by deep fluids, we combine the metamorphic basement with these two types of reservoirs to compare parameters (gas abundance, porosity, and permeability).

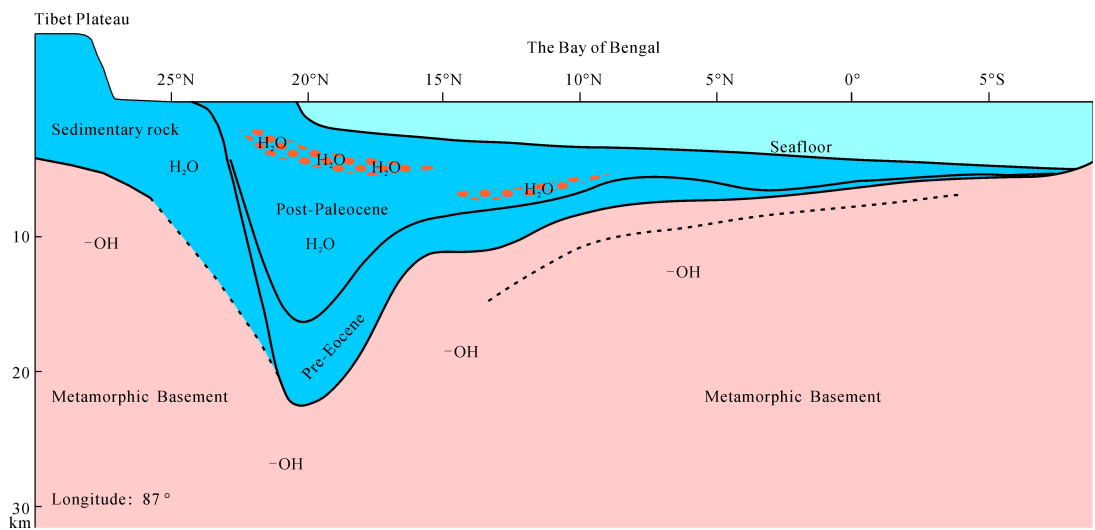

**Figure 2.** Sedimentary strata and metamorphic basement (edited from Curray, 1991).

Let us first look at some parameters of shale. Taking the Fuling shale gas field as an example, the abundance of adsorbed gas in carbonaceous shale from well Jiaoye 1 (JY1) is 0.5-2.5 $m^3$/t (Li et al., 2014). Liu et al. (2017) analyzed the physical characteristics of core samples from shale in the Longmaxi Formation in the Changning and Weiyuan regions and found that the core porosity values range from 1% to 10%, with typical values of 6%.

Deep mantle fluids take the form of structural water in fluid inclusions or mantle minerals. Among water-rich mantle minerals and regions, pyroxene contains water at concentrations of

~200–500 ppm (Bell & Rossman, 1992); these quantities are equivalent to ~5%–10% of the aforementioned maximum shale gas reservoir capacity. These abundant deep fluids originate mainly from the rocks in the metamorphic basement whose porosity is relatively small (Harms, 1994; Huenges et al., 1997).

In summary, the porosities, permeabilities, and deep fluid abundances of crystalline metamorphic bedrock are all much lower than the corresponding values for shale (Table 1). Shale releases the accumulated gases only when the shale is fractured. Therefore, deep fluids are unlikely to rapidly escape from the rock lattice and release the energy required to produce an earthquake.

**Table 1**. Comparison of parameters between deep fluids and oil and gas.

|  | Metamorphic bedrock | Shale | Conventional reservoir |
| --- | --- | --- | --- |
| Location | crystalline basement | basin | basin |
| Porosity | <1% | 1-5% | 5-30% |
| Permeability | <0.1 mD | 0.1-50 mD | 50-5000 mD |
| Limit Drainage Radius | 1 nm | 1 m | 500 m |
| Abundance of fluid/gas | <0.05 $cm^3/g$ | 7-15 $cm^3/g$ | >8000 $cm^3/g$ |

## 3 Methodology and Model – the Possibility of a Physical Explosion

The lithosphere can be regarded as a two-phase medium comprising solid and fluid. Since the energy directly generated by a fault dislocation, that is, due to the activity of the solid component, is not sufficient, this paper proposes that the main energy of natural earthquakes may be contributed by shallow fluid: a fault caused by tectonic movement fractures a sedimentary rock with high porosity, which causes the high-pressure fluid in the rock to release energy suddenly, thereby enhancing the destructiveness of the earthquake.

High-pressure fluids, like solids, expand and release elastic energy when relieved of their pressure, which is similar to a physical explosion. In marine seismic exploration, an air gun is used as the vibration source to generate seismic waves (Lv et al., 2020). Wang et al. (2012) placed four large air tanks in a lake and released the pressure in them at the same time, which could generate an earthquake of magnitude 0.5. Zhang et al. (2018) used liquid carbon dioxide under high pressure to blast rock instead of ordinary explosives. All of these examples are phenomena in which the fluid or gas pressure is released.

Unfortunately, at present, the general belief is that the strong waveforms after first motions are caused by S-waves, while this kind of pressure release or fluid explosion can produce only P-waves. Seismologists generally use P/S-type spectral ratios of regional phases (e.g., Pg/Lg, Pn/Lg, Pn/Sn) to determine that an earthquake was not caused by an explosion event (Zhang & Wen, 2015; Zhao et al., 2014).

Therefore, some questions need to be answered: Will energy be released when an earthquake fault penetrates a porous reservoir? Is there voluminous fluid in the reservoir? How much energy can be released? Finally, how different is the energy release signal from an earthquake signal?

3.1 The release of fluid pressure in the reservoir

For common explosion phenomena, the concentration of explosives is large, and almost all of these explosives participate in the event. Under normal temperature and pressure conditions, the pores of reservoir rock are not connected, and the fluid within the pores cannot flow out spontaneously; thus, suggesting that fluid can produce an explosion is indeed difficult to understand. However, the blowout phenomenon has confirmed that the fluid pressure within a certain width in the reservoir can be released; this process is also a safety accident that needs to be prevented at all times during drilling (Pinkston & Flemings, 2019; Tingay et al., 2008).

Reservoirs are under a high pressure, which is characterized by the pressure coefficient. The pressure coefficient of a reservoir is greater than 1.0 in most cases, such as 1.8 times in the West Sichuan Depression (Leng et al., 2011), and the pressure can reach 72 MPa at a depth of 4 km (Figure 3a). Then, sand needs to be mixed into the water to increase the drilling fluid density. Sometimes, the reservoir fluid pressure is underestimated before drilling; for example, the drilling fluid density may be only 1.5 g/cm$^3$, and the fluid pressure in the wellbore at a depth of 4 km can be 60 MPa (Figure 3a). Compared with the pressure in the reservoir at a depth of 4 km (72 MPa), the pressure produces a surplus differential stress $\Delta P$ of 12 MPa. In this case, blowouts will occur, and the drilling rig will be flushed, causing major safety accidents. At the moment of fault dislocation, when the fault zone is not full of water, it is a vacuum, and the pressure difference of the pore fluid is then 72 MPa relative to the fault zone. This is the mechanism of pore fluid pressure release. The fluid within a certain distance r from the wellbore can flow freely into the wellbore (as shown by the yellow area in Figure 3b), while the fluid outside this range cannot and remains at high pressure. This distance is called the well spacing or limit drainage radius (LDR) in petroleum exploration (Li, Zhou, et al., 2017). Water flooding technology (Tetteh et al., 2021) injects high-pressure water with a pressure of $\Delta P$ into the wellbore at the wellhead of the water injection well (Figure 3c). If the distance between the two wells is greater than r, the driving process fails. Yuan et al. (2007) concluded that the well spacing r can be designed as 1000 -1500 m when the permeability is 5-150 mD. For low-permeability reservoirs, the r value is approximately 250 m.

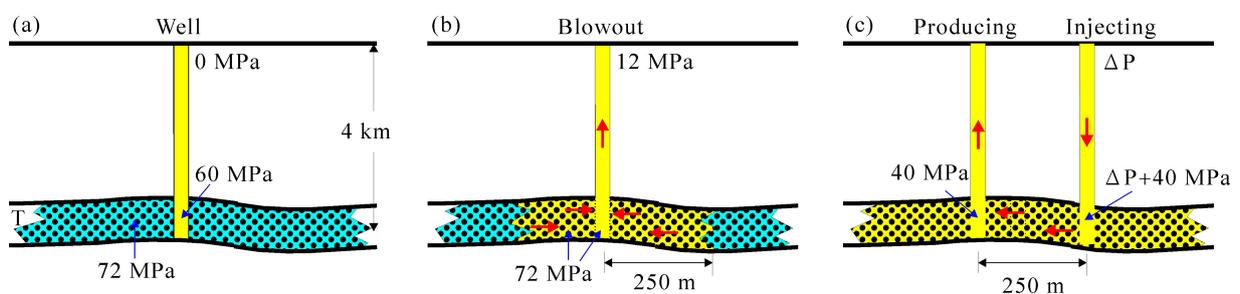

**Figure 3**. Variations in reservoir and wellbore pressure before and after drilling through the reservoir. (**a**) Before perforation; (**b**) during perforation; (**c**) water flooding mode.

3.2 The amount of fluids in sedimentary formations

In all kinds of rocks, only some sedimentary rocks contain large pores and are rich in water (Table 1). There are sedimentary strata at the epicenter of the 2008 Wenchuan earthquake, which has a tectonic and sedimentary background similar to that of the Sichuan Basin. Scientific drilling of WFSD-1 after the earthquake confirmed that the Triassic Xujiahe Formation, which is also the main oil- and gas-producing layer in the Sichuan Basin, was present at a depth of 1000 m (Fang et al., 2020). Therefore, it would be more accurate to say that the Sichuan Basin should contain the epicenters of earthquakes. Therefore, this paper uses the reservoir scale of the western Sichuan depression to explain the amount of fluid involved in the Wenchuan earthquake vertically and laterally.

Vertically, the cumulative thickness of high-porosity sandstone in the Xujiahe Formation is more than 500 m (Zhao et al., 2013). At the same time, many sets of high-porosity reservoirs, such as Silurian, Ordovician and Sinian reservoirs, have developed. The well-known reservoirs are the Triassic Xujiahe Formation and Sinian Dengying Formation, with porosities of 10%-30% (Table 2). From the actual oil and gas exploration profile, there are many layers with high porosity and a large cumulative thickness, and these features are unique to sedimentary formations. As shown in Figure 4, the red area in the profile represents the high-porosity reservoir (Wu et al., 2015). On May 4, 2020, drilling was carried out in Tianbao town, and a daily output of 1,219,800 m$^3$ of natural gas was obtained from the Sinian Dengying Formation.

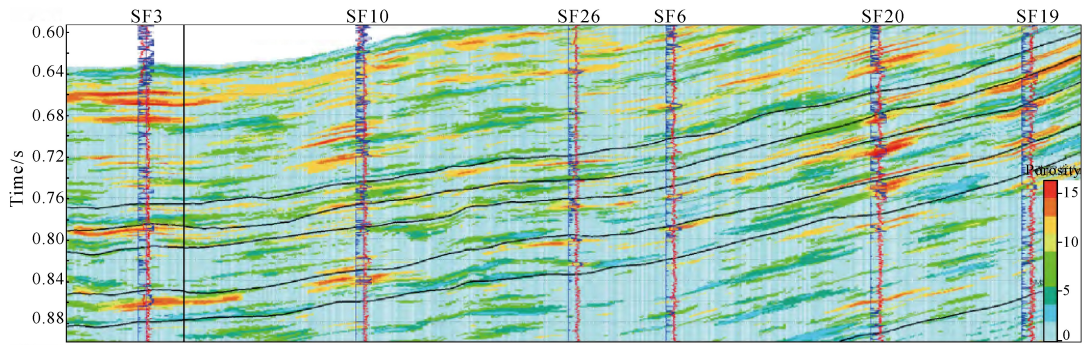

**Figure 4**. Cross-well porosity section from geostatistics modeling of the reservoirs in the Shifang area, western Sichuan depression.

**Table 2**. Comparison of reservoir parameters and model parameters in the western Sichuan depression.

| Reservoir parameters | Measured parameters | Model parameters | Data source |
|---|---|---|---|
| Length | 10-50 km | 5 km | |
| Width | 5-22 km | 0.5 km | |
| Thickness | 150-300 m | 100 m | Zhao et al. (2013) |
| Porosity | 5%-20% | 5% | |
| Depth | > 6 km | 4 km | |
| Pressure of fluid | > 108 MPa | 72 MPa | Leng et al. (2011) |
| Abundance of fluid | The proven reserves of Yazihe gas field are > 30.9×10$^6$ m$^3$ (to 2014) | 13×10$^6$ m$^3$ | |

Laterally, several large northwest-southeast-oriented traps have been found in the western Sichuan depression closest to the Wenchuan earthquake. There is a natural gas field only 50 km from the epicenter (Figure 5a), namely, the Duck River gas field, with a length of 60 km and a width of 5-20 km (Ma et al., 2019). These large traps use the Triassic Xujiahe Formation as a reservoir with a fluid volume of approximately 375 million to 66 billion cubic meters. On May 12, 2020, a well was drilled in the southern section of the Longmen Mountain structural belt close to the epicenter, and 668,600 m³ of gas was obtained daily from the lower Permian Qixia Formation. The Yingxiu epicenter can also be speculated to have voluminous fluid.

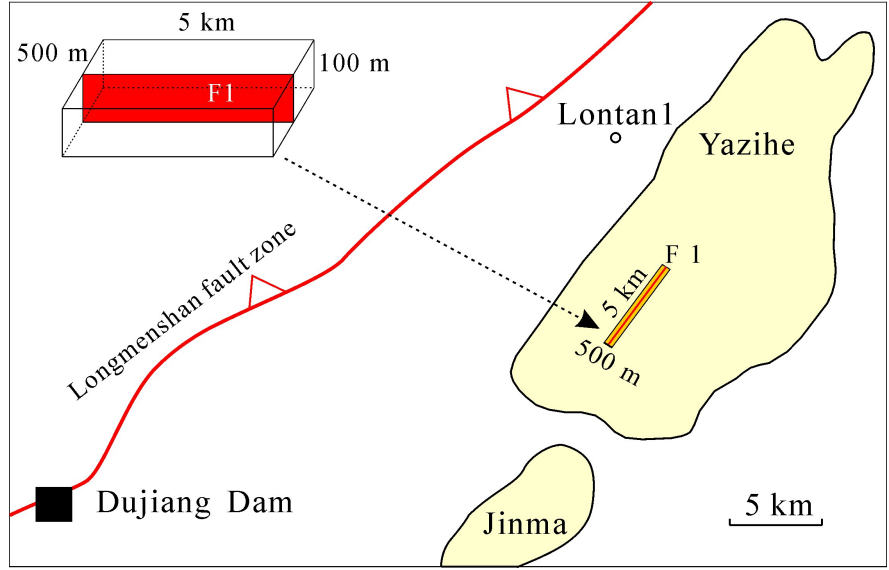

**Figure 5**. Schematic diagram of the Yazihe gas field and fracture model range.

3.3 Energy released by reservoir fluid

A blowout is limited to a very small amount of participating fluid, and its energy is small; if a seismogenic fault caused by tectonic movement fractures a long reservoir, it may release much more energy than a blowout.

Therefore, we designed a reservoir with a length of 5 km, a width of 500 m, and a thickness of 100 m, with a burial depth of 4 km, as shown in Figure 5b, and compared it with the Yazihe gas field on the same map (Figure 5a). The contrast map illustrates that it is very easy for a fault to break open the 5 km-long reservoir model; the width is calculated according to the LDR of 250 m on both sides of the fault. Assuming a reservoir porosity of 5%, the volume of the pore fluid in our model is approximately $13 \times 10^6$ m³. When fault $F_1$ fractures the reservoir, these fluids are released from a pressure of 72 MPa to a hydrostatic pressure of 40 MPa corresponding to a depth of 4 km, and the amount of volumetric expansion will reach approximately 1.4%. For the released energy, we adopt the pressure release formula of water as follows:

$$E_L = \frac{(\Delta P - 1)^2 V \beta_t}{2} \times 10^8 \tag{1}$$

where $E_L$ is the energy released when the liquid pressure vessel explodes at normal temperature (kJ), $\Delta P$ is the pressure change of the liquid (MPa), V is the volume of the reservoir (m$^3$), and $\beta_t$ is the compressibility of the liquid at pressure P and temperature T (let $\beta_t = 4.4 \times 10^{-4}$ MPa$^{-1}$). We calculate that the energy released is:

$$E_L = (32-1)^2 \times 0.138 \times 10^8 \times 4.4 \times 10^{-4}/2 \times 10^8 \times 1000 = 2.748 \times 10^{17} \text{ J} \qquad (2)$$

This energy is equivalent to that of a magnitude 8.4 earthquake.

The setting of the above parameters is not strict compared with the data obtained from the actual exploration, and only the lower limit value of the parameters obtained by the oil and gas exploration results is taken (Table 2). For example, the measured value of reservoir porosity is 5% - 20%; the measured reservoir pressure coefficient is 1.8 to 2.0. In addition, at the moment when the fault is just dislocated, the pressure in the fault zone is 0, the fluid pressure in the reservoir should directly decrease from 72 MPa to 0 MPa, and the amount of volumetric expansion will reach approximately 3.2%, so the energy would be greater. Therefore, the above energy estimation is reasonable.

**4 Release Mode of Fluid Pressure and Analysis of the Wenchuan Earthquake Signal**

4.1 The mode of fluid pressure release

Based on the geological profile (Figure 4) of the Shifang area in the western Sichuan depression, a model of the elastic energy release from a high-pressure fluid is proposed. Figure 6 shows the physical explosion pattern; Figure 6a indicates that before the earthquake, the Triassic reservoir (T) was not disturbed; once the fracture caused by tectonic movement pierced the reservoir, the fluid in the reservoir quickly entered the fault zone, forming a high-pressure fluid capsule (yellow area in Figure 6b) and generating foreshock vibrations. When the fault reached the surface, the high-pressure fluid capsule released the overpressure $\Delta P$ and produced strong earthquakes. This high-pressure fluid capsule filled the whole fault zone, so the vibrations were generated not from a point source at the moment of pressure release but from a surface/body source. The depth of the hypocenter is not the depth of the shallow reservoir but the equivalent depth of the deeper high-pressure fluid capsule. Therefore, the actual evolution of an earthquake can be divided into two processes: fault rupture (represented by II) and fluid pressure release (represented by III). Before the initial motion, there is a microfracture process (represented by I), for which the vibration signal is not received, but the electrokinetic effect produced by the fluid activity can be detected. Fault rupture (II) can be described by modern earthquake theory, which is accepted in this paper, and can produce weak amplitudes before the mainshock; only when a fracture develops gradually and penetrates the high-pressure reservoir can a strong earthquake occur.

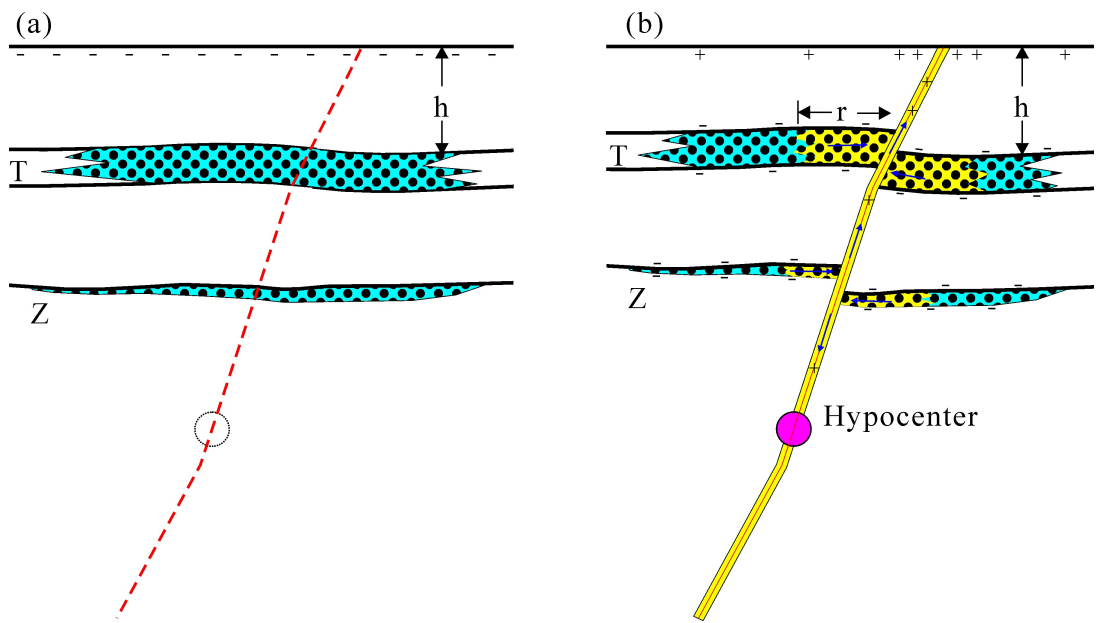

**Figure 6**. Profile of physical explosion mode. (**a**) Unruptured state; (**b**) post-fracture state.

4.2 Direct evidence

The following paragraphs describe several aspects of direct evidence for physical explosion.

First, we can see from seismic signal analysis that the seismic process can be divided into two categories, namely, the fault rupture (II) and the strong earthquake (III), and the strong waveforms are caused by events in different places. Figure 7 presents the seismic waveform from the 2008 Wenchuan Ms 7.9 earthquake, from which we can see the two processes. The connection line of the initial movement of each station is A, and that of each station before the mainshock of the earthquake is B. The direct P-wave Pg between A and B has a low amplitude, indicating the fault rupture process (II), and the right side of B represents the second process, namely, multiple strong waveforms caused by the subsequent multiple events of fluid pressure release (III), which is generally considered to be the direct S-wave Sg. Line A is not parallel to line B, and the slope of line A is small, which indicates that the fracture is developing in the same direction as the wave propagation. The arrival time interval between the initial motion A and the high-magnitude earthquake B at Wolong station in the first curve is only 1.1 s, while the arrival time interval between the initial motion and the high-magnitude earthquake at Chonghua station in the fifth wave curve is approximately 25 s. This time duration represents the total fracture time of the fault, which is consistent with previous estimates of 22 s (Shang et al., 2015). Although the Qinpin station is far from the epicenter (90 km), it is close to the fault zone. Therefore, the maximum frequency of its initial motion is 60-70 Hz (H1'), which is basically the same as that at Wolong station (H1), although the strong waveform of S1 has a certain attenuation (A1'< A1) (Figure 8). Compared with Wolong station, Qinpin station is closer to the location of event S2, with a strong amplitude (A2'> A2) and a wide frequency band, and the highest frequency at Qinpin (H2') is higher than that at Wolong station (H2).

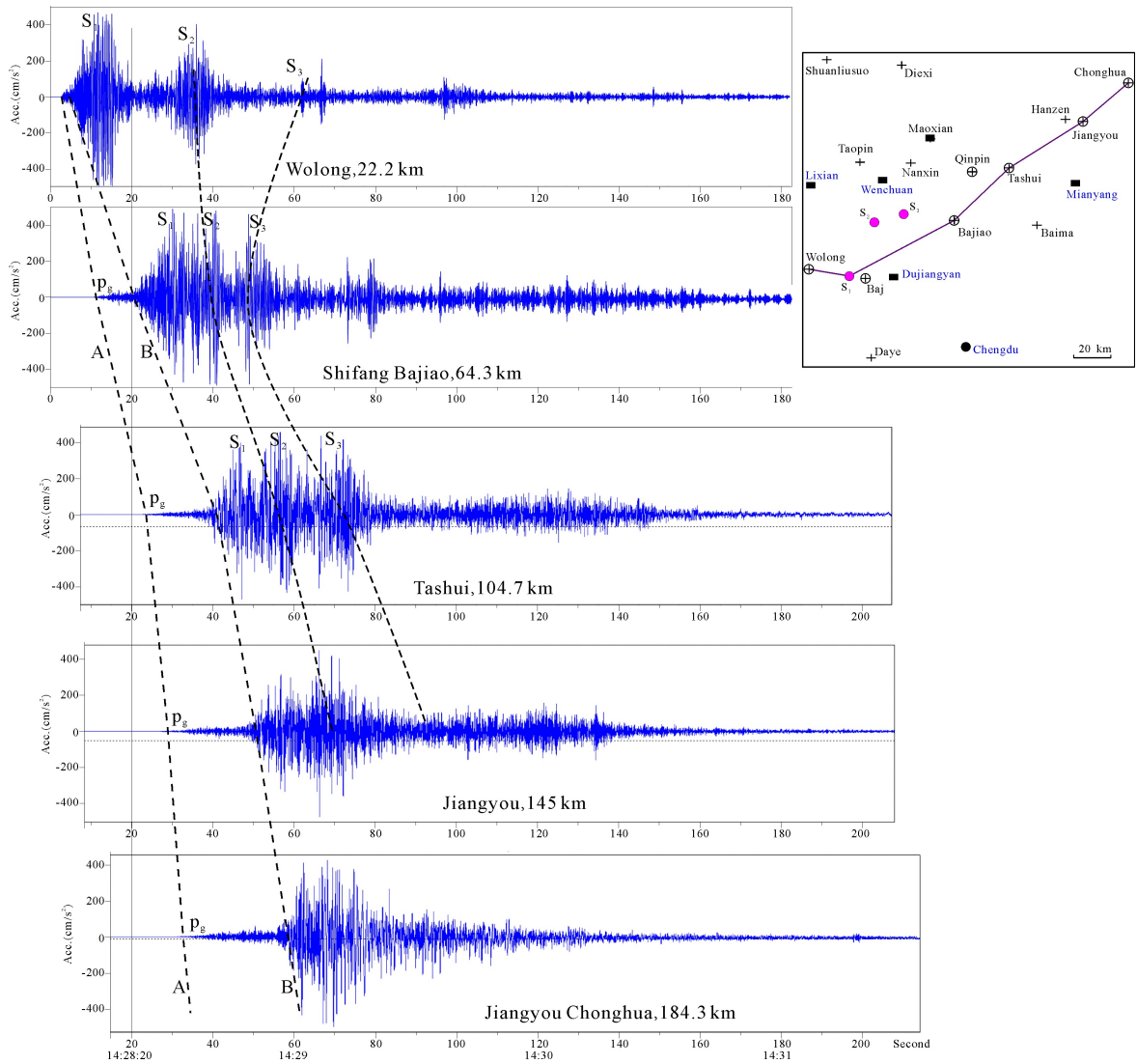

**Figure 7**. Comparison of seismic waveforms from the Wenchuan earthquake at various stations.

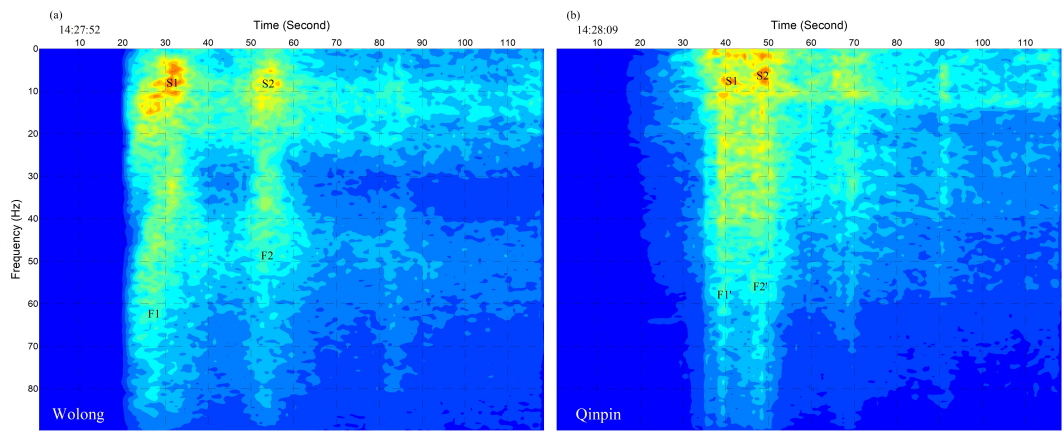

**Figure 8.** Time frequency analysis at seismic stations near the Wenchuan earthquake. (**a**) Time frequency diagram at Wolong station; (**b**) time frequency diagram at Qinpin station.

Second, from the propagation path of seismic waves, the strong waveform is the P-wave, which is not currently recognized by seismologists because general seismic phase analysis is based mainly on the study of moderate earthquakes, distant earthquakes and initial motions, while less attention is paid to the strong waveforms observed near the epicenter. According to the three components (X, Y, Z) of the seismic wave received by the Wolong seismic station nearest to the epicenter (18.7 km) (Figure 9a), the Z component (blue line) dominates the seismic wave between 14:28:2.6 (initial motion) and 14:28:14.0, and the signal of Z then weakens and is gradually dominated by the X component with increasing time (Figure 9a). The initial motion waveform with a duration of 1.4 s at Baj west of the Zipingpu Reservoir, only 5 km from the epicenter, shows a similar pattern (Ye et al., 2008). The polarization direction of $p_{g2}$-$p_{g5}$ is basically the same as that of initial motion $P_{g1}$, and the X direction component of $P_{g6}$ is greater than the Z component, indicating that the event is from the epicenter of the Wenchuan earthquake, which is farther north than Yingxiu. Considering that the low-velocity zone on the surface causes ray deflection (Figure 9b), only the P-wave is stimulated at hypocenter H, which can conform to the law. If a shear wave is stimulated at the hypocenter, the vibration direction of the particle should be close to the horizontal direction when the seismic wave reaches the Wolong station; that is, the received seismic wave should be mainly in the X direction. Therefore, the strong waveform (Figure 9a after 3.7 s or line B in Figure 7) is not the direct S-wave or the surface wave but another series of events produced by the event of fluid pressure release at another place H after the initial rupture A.

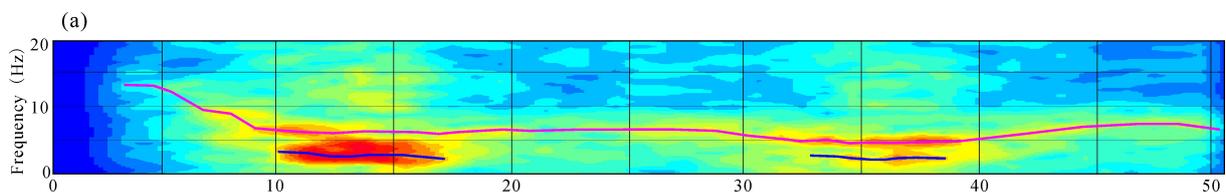

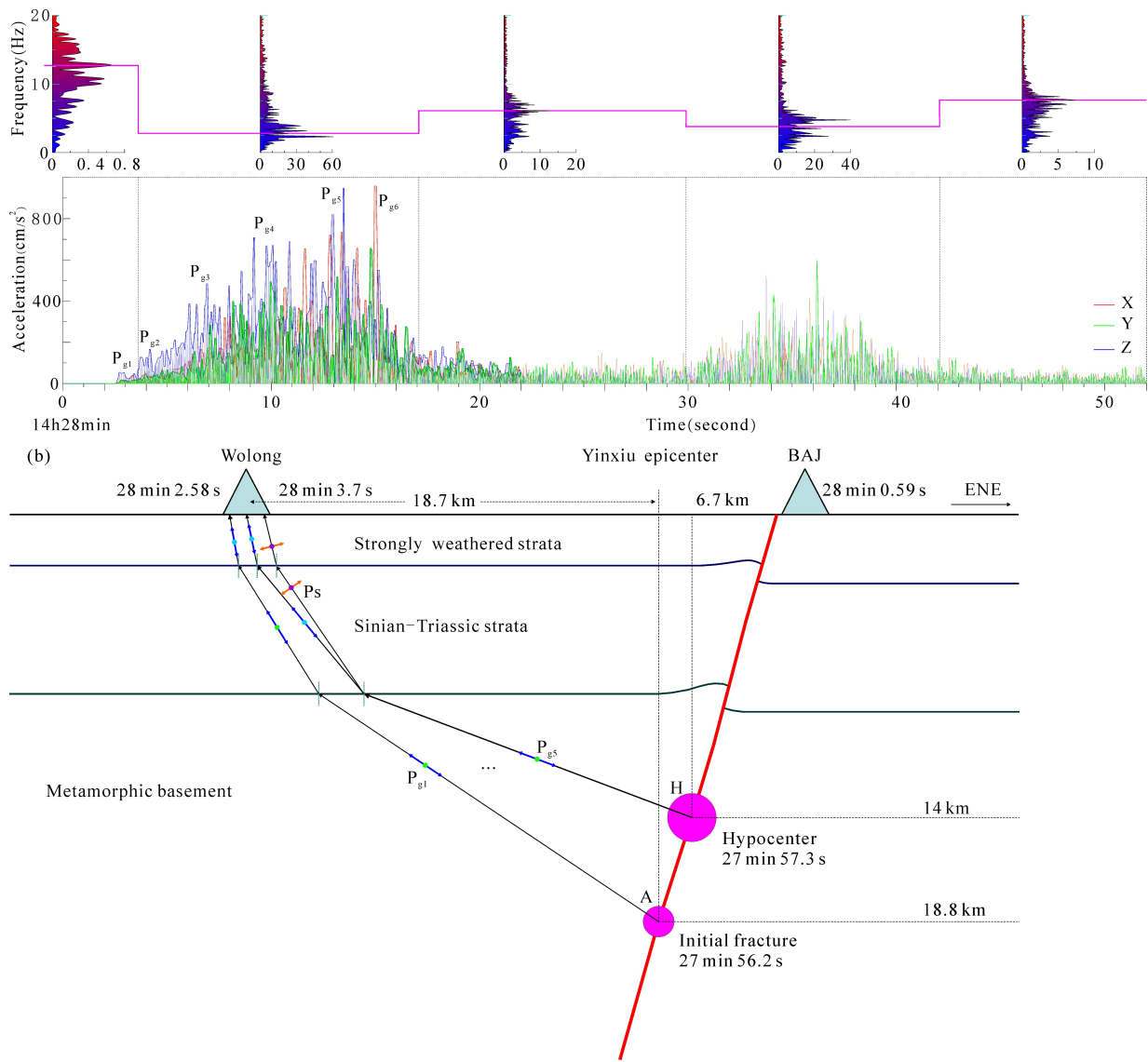

**Figure 9**. Amplitude envelope and seismic ray path of seismic signal received by Wolong seismic station. (**a**) The energy envelope of the three components and time-frequency analysis of the X component; (**b**) schematic diagram of seismic wave propagation path.

Some scholars use the rule that the displacements of the P-wave and S-wave at far-field seismic stations are inversely proportional to the square of propagation velocity to explain why the energy of the S-wave is stronger than that of the P-wave and the S-wave is the mainshock. The premise is that P-waves and S-waves of the same strength can be produced at the hypocenter.

Third, the seismic waves show dual basic frequency characteristics.

We used the fast Fourier transform (FFT) to calculate the amplitude spectrum of seismic signals in X direction of Wolong station at different time periods(the first curve in Figure 7) and found that the dominant frequencies of waveforms with low-amplitude wave signals in 3 time periods of 2-14.7 s, 26-38 s, and 54-62 s are 12 Hz, 6 Hz, 7.5 Hz, respectively (Figure 9). The dominant frequencies of the strong waveforms in the 2 periods from 14.7-26 s and 42-50 s are 2.3 Hz and 4.8 Hz, respectively. The dominant frequencies for 3.7-16 s and 32-40 s are 2.3 Hz and 2.5 Hz, respectively; i.e., these are strong-amplitude fluctuations. The continuous time-frequency analysis diagram (Figure 9a) shows that the two strong waveforms obviously contain another high-frequency component that is basically consistent with the initial rupture and that the frequencies are 6 Hz and 5 Hz; this diagram further indicates the superposition of two kinds of waves. That is, fault dislocation produces weak signals with high frequency like those from the hydraulic fracturing of shale, while the vibration caused by the release of fluid pressure is an area vibration source with a large amount of fluid and high energy but low frequency. Similar to the seismic wave produced by a dynamite explosion, the frequency is inversely proportional to the amount of explosive:

$$\frac{1}{f} = k_2 Q^{m_2} \tag{3}$$

where $f$ is the frequency, $Q$ is the explosive quantity and $k_2$ and $m_2$ are constants. The larger the hypocenter body is, the easier the production of low-frequency seismic wavelets. This dual basic frequency characteristic is consistent with the time-frequency analysis of Gao et al. (2020).

Fourth, strong traces of fluid activity were found in drilling the fault zone. Wang et al. (2015) observed core samples from borehole No. 1 (WFSD) of the Wenchuan earthquake fault scientific drilling project by using scanning electron microscopy (SEM) and transmission electron microscopy (TEM); these authors found that much fluid may flow into the fault zone during the coseismic process, as shown in Figure 10a.

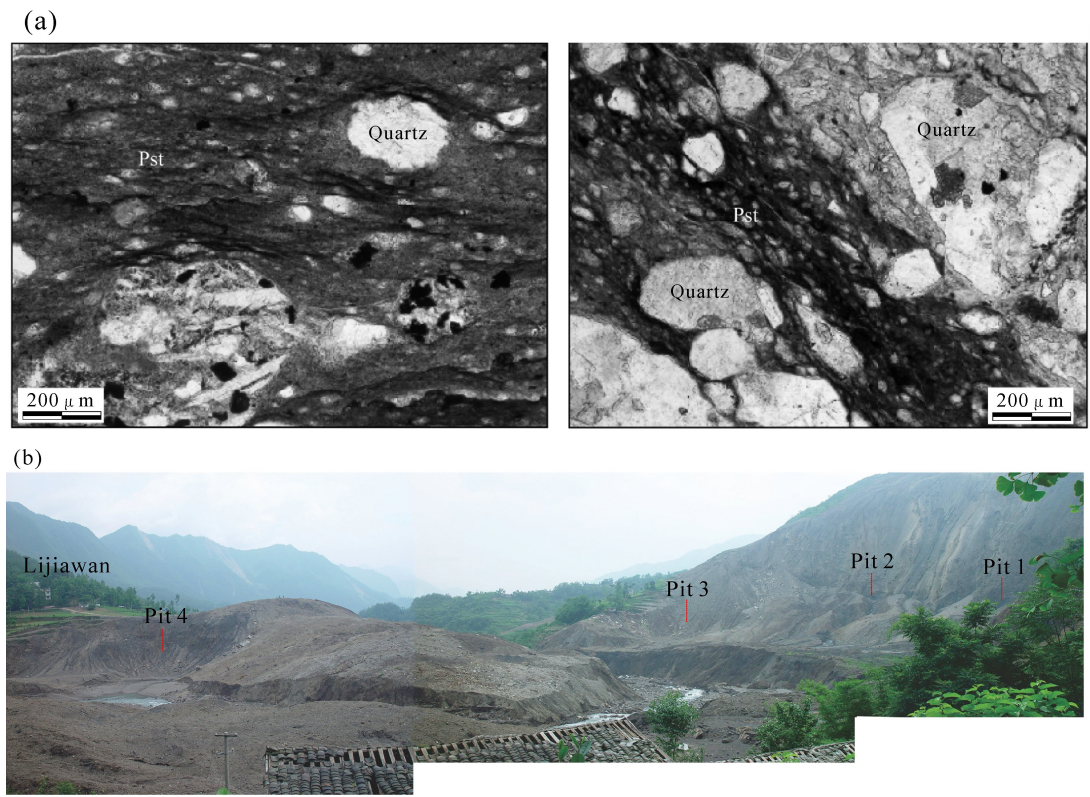

**Figure 10**. Fluid activity and explosion during Wenchuan earthquake. (**a**) Flow structure in pseudotachylyte (Pst) from the Bajiaomiao outcrop; (**b**) four explosion pits on the surface of Shuijingyan landslide and weir plug body in Guixi town, Beichuan (photography date: June 20, 2008; lens direction: S, 31.973°N, 104.603°E).

Fluid activity plays a role not only in the strong earthquake stage but also before the initial motion, which has attracted more and more attention in recent years; this may be a major breakthrough for earthquake prediction in the future. Li, Zhang, et al. (2017) reported that atmospheric electric field anomalies were observed at 58 km from the epicenter of the Wenchuan earthquake. Varotsos and Alexopoulos (1984) used geoelectric field observational data to explore precursor seismic electric signals (SES). Hao et al. (1998) counted the atmospheric electric fields of 21 earthquakes, all of which produced strong negative anomalies (- 400 ~ - 936 V/m) from 3-38 days before the initial motion. Korsunova et al. (2013) and other scholars also observed similar phenomena. Chen et al. (2021) studied the characteristics and influencing factors of negative anomalies and established a monitoring network for impending earthquake prediction.

Some scholars believe that the electric field disturbance is produced by piezoelectric effects (Finkelstein & Powell, 1970) or by stretching of the sedimentary rocks (Marapulets & Rulenko, 2019), but the electric field generated by fast fluid flow in porous rocks is much larger than those generated by the other effects. Finkelstein et al. (1973) denied the theory that the piezoelectric effect leads to earthquake lightning (EQL) by experiment. The streaming potential was generated when the fluid was flowing, and its magnitude was proportional to the pressure difference driving the fluid flow (Fitterman, 1978). The streaming potential effect produced by water injection in the process of oil production has been recognized for many years and can be observed at the surface (Wurmstich & Morgan, 1994). There exists a double layer between the reservoir skeleton and the fluid (Lorne et al., 1999), which can be destroyed by vibration or fluid flow and generate an electric current. In a quiet period between earthquakes, on a sunny day with no wind-blown sand, the surface generally presents a negative charge (Figure 6a), so the atmospheric electric field is positive. Once a fault ruptures, the multiple reservoirs will be connected vertically; fluid with a certain degree of positive charge will be ejected upward to the shallow layer at high speed due to the large pressure difference between the fluids at different depths. As this positive charge approaches the surface, it counteracts the negative charge of the surface and gives the surface a positive charge (Figure 6b); thus, we can observe the reversal of the atmospheric electric field, which becomes negative as described above. This transient change in the potential difference caused by the abruptness and randomness of rupture is also confirmed from the observed signal (Varotsos & Alexopoulos, 1984). The negative atmospheric electric field before the initial motion is caused by small-scale fluid activity produced by microfractures in the seismogenic stage (I), in which there is no obvious fault dislocation and the vibration signal is too weak to be detected. During hydraulic fracturing, although fractures with a height of approximately 200 m can be generated in the stratum at depths of 3-5 km, the effective vibration signal cannot be detected 2-5 km away from the surface. Certainly, the atmospheric electric field anomaly will continue after the initial motion, but the electrokinetic effect is more intense, such that even the strong discharge phenomenon of earthquake lightning appears (Fidani, 2010; Finkelstein et al., 1973; Kamogawa et al., 2005). Clearly, only sedimentary strata host fluid activity.

Fifth, earthquake witnesses saw the explosion. Based on reports from those who experienced the Wenchuan earthquake, this earthquake event, which occurred in three stages, was a continuous explosion that lasted for two minutes. The sound of the explosion was dull, and explosion debris reached heights of up to 50 m. These explosions were not caused by urban natural gas pipelines or other human factors. In fact, in the Longmenshan fault zone, which is approximately 200 km long, at least five large explosion sites can be clearly observed. Figure 10b shows one of them. There are four black smoke pits oriented from west to the east in the picture; one pit corresponds to one blasting position, and clean black stones were spewed out, accompanied by heat and air waves (Shang et al., 2015). These observations are consistent with the seismic signals shown in Figure 7.

Sixth, the total rupture length of the Wenchuan earthquake reached 200 km, but the rupture involved in a single-source body is not long. According to the stress drop caused by a fault dislocation, the elastic energy released cannot reach a very high magnitude and must be due to other factors. The distances between the multiple sources of the Wenchuan earthquake are only 5-20 km, as speculated by our predecessors. Obviously, the elastic energy released by a fault dislocation cannot be used to explain the tremendous energy of the Wenchuan earthquake.

4.3 Indirect evidence and some earthquake examples

First, cryptoexplosive breccias developed all around the world indicate the universality of explosion phenomena (He & Qiao, 2015; Pope et al., 1997; van Loon et al., 2016).

Second, the lag between the initial motion and the strong earthquake can be observed in many seismic events. The 2009 L'Aquila Ms 6.2 earthquake in Italy actually lasted 7 days between the initial motion and the mainshock (Marzocchi et al., 2014). The lag for the 2001 Kunlun Mountain Ms 8.1 earthquake was three days (Hu, 2018). However, Hu inferred that the precursor wave of the earthquake was caused by an extreme windstorm in the North Atlantic, and this type of earthquake is considered to be a slow earthquake and to be caused by fault creep (Beroza & Jordan, 1990; Gao & Wang, 2017). In fact, the maximum magnitude of the precursor wave of this earthquake was equivalent to approximately Ms 3.5 seismic activity, and its focal depth of 12 km was likely to have been located in the sedimentary layer, which was conducive to physical explosions, but the fault had not penetrated the reservoir before the mainshock. In 2018, the Qiangcan-1 well, located in Shuanghu County, northern Qiangtang, on the Qinghai-Tibet Plateau, was drilled to a depth of 4,000 m and did not penetrate the Permian strata.

Third, the major earthquakes in the world occur in areas with sedimentary strata, such as earthquakes in the circum-Pacific seismic belt located on the continental shelf or slope, and shallow earthquakes above Ms 6.0 within 30 km depth account for approximately 65% (2012 to present). The epicenter of the 2011 earthquake on the Pacific coast of Tōhoku (38.1°N, 142.6°E) was located in the interior of a basin, specifically in the southern part of the Kitakami basin (Arai et al., 2014), which is rich in oil and gas. The 1960 Valdivia (Chile) Mw 9.5 earthquake, the largest earthquake in the world, occurred on the continental shelf close to mainland Chile, where the slab and the overlying mantle wedge show local decreases in seismic velocities possibly caused by hydration or underplating of sediments (Dzierma et al., 2012). Two subsequent major earthquakes to the north of this epicenter, namely, the 2010 Maule Mw 8.8 earthquake and the 2015 Illapel Mw 8.3 earthquake (Ruiz et al., 2016), were also located in similar positions on the continental shelf of Chile. Olsen et al. (2020) confirmed that the incoming sediments along the south-central Chile margin where these earthquakes occurred are composed almost entirely of trench wedge turbidites with a thickness of more than 8 km according to multiple high-precision seismic profiles. These earthquakes are usually considered results of plate subduction, but the 1976 Tangshan Ms 7.8 earthquake cannot be explained by plate movement. Deep below the epicenter of the earthquake are Ordovician and Cambrian limestone strata (Liu et al., 2011) containing high-porosity and high-permeability reservoirs; thus, this earthquake can be well explained by the view of fluid pressure release.

Fourth, the blowout phenomenon can be explained by the pressure release of the fluid in the reservoir. The Lusi mud eruption was not thought to have been triggered by the Yogyakarta Ms 6.3 earthquake but by a nearby blowout in well BJP-1 (Tingay et al., 2008). In fact, the BJP-1 drilling results show that there is a clay cap at depths of 900-1,800 m directly covering high-porosity volcaniclastic sand with a thickness of 1000 m containing high-pressure water, which meets the fluid pressure release conditions described in this paper. Earthquakes, blowouts and mud eruptions are all different manifestations after puncturing high-pressure reservoirs.

Fifth, based on our model, we infer that the shortest distance for another strong earthquake to occur in the same area is the LDR (approximately 250 m), because reservoirs farther from the original fault than the LDR will maintain high fluid pressure and a strong earthquake will still develop once rupture occurs. For example, there were 11 earthquakes in the Songpan area of Sichuan Province within the scope of 16 * 53 km$^2$ in 18 days from August 16 to

September 2, 1976 (104.3°, 32.5°) (Jones et al., 1984). The two largest earthquakes with magnitude 7.2 were separated by 7 days, and the epicentral distance of two pairs of earthquakes separated by one week was only 5 km.

**5 Discussion and Conclusions**

The following conclusions are drawn by combining information from seismology, geochemistry, geology, and drilling:

(1) Deep fluids are unlikely to cause an earthquake.

(2) The pressure release of high-pressure fluid in sedimentary strata may be an important energy source for destructive strong earthquakes. When a fault caused by natural tectonic movements fractures a reservoir, the elastic energy of the high-pressure fluids in the reservoir can be released, and the released energy can exceed the energy released by an earthquake of magnitude Ms 8.0. Human production activities, such as drilling, fracturing, and water injection into wells, can also break into a reservoir, and the pressure of the fluid in the reservoir can be released, which causes perceptible induced earthquakes. The essence of an earthquake is equivalent to piercing a "high-pressure chamber" within sedimentary strata.

(3) From an analysis of the near-earthquake signal of the Wenchuan earthquake in the time and frequency domains, we can see that the strong waveform signal comes from the vibration of the P-wave, not the S-wave.

(4) According to the characteristics of the reservoir, the shortest distance for another earthquake to occur in the same area should be greater than the LDR (>250 m).

The scientific problem to be discussed is how to use the atmospheric electric field disturbance caused by the electrokinetic effect of flow motion to quantitatively evaluate the locations and depths of microfractures and thus to carry out earthquake prediction.

5.1 Data and resources

The earthquake data came from the Seismological Bureau of Sichuan Province with their consent. Other data came from published papers.

**Acknowledgments, Samples, and Data**

We would like to thank the staff of the Sichuan Seismological Bureau for providing the waveform data of the Wenchuan earthquake, as well as Professor Zhao Yong and Huang Fuqiong from the State Seismological Bureau, Professor Liu Yufa from the Sichuan Seismological Bureau and Professor Chen Tao from the National Center for Space Sciences of China.

The authors declare no competing interests.

**Data Availability Statement**

The raw/processed data required to reproduce these findings cannot be shared at this time as the data also forms part of an ongoing study.